\documentclass[twocolumn,aps,pre,showpacs,superscriptaddress,nofootinbib]{revtex4}

\usepackage[latin1]{inputenc}
\usepackage{graphicx,epstopdf}
\usepackage{amsmath}
\usepackage{xcolor}
\usepackage{hyperref}
\usepackage{amsfonts}
\usepackage{amssymb}
\usepackage{color}
\usepackage{latexsym}
\usepackage{times,txfonts}
\usepackage{braket}
\usepackage{textcomp}
\usepackage{float}
\hypersetup{
    colorlinks=true,
    linkcolor=blue,    % Color de las etiquetas de secciones, ecuaciones, etc.
    urlcolor=blue,     % Color de los enlaces URL
    citecolor=blue     % Color de las referencias bibliográficas
}

\newcommand{\fig}[1]{{Fig.}}

\begin{document}
   
\title{Qubit dynamics of ergotropy and environment-induced work}

%%%%%%%%%%%%%%%%%%%%%%%%%%%%%%%%%%%%%%%%%%%%%%%
%%%%%%%%%%%%%%---Authors---%%%%%%%%%%%%%%%%%%%%
%%%%%%%%%%%%%%%%%%%%%%%%%%%%%%%%%%%%%%%%%%%%%%%
\author{J. M. Z. Choquehuanca}
\affiliation{Instituto de F\'isica, Universidade Federal Fluminense, Av. Gal. Milton Tavares de Souza s/n, Gragoat\'a, 24210-346, Niter\'oi, RJ, Brazil}

\author{P. A. C. Obando}
\affiliation{School of Physics, University of the Witwatersrand, 1 Jan Smuts Avenue, Braamfontein, 2000, South Africa}

\author{F. M. de Paula}
\affiliation{Centro de Ci\^ encias Naturais e Humanas, Universidade Federal do ABC, Avenida dos Estados 5001, 09210-580, Santo Andr\'e, S\~ao Paulo, Brazil}

\author{M. S. Sarandy}
\affiliation{Instituto de F\'isica, Universidade Federal Fluminense, Av. Gal. Milton Tavares de Souza s/n, Gragoat\'a, 24210-346, Niter\'oi, RJ, Brazil}

\date{\today}

%%%%%%%%%%%%%%%%%%%%%%%%%%%%%%%%%%%%%%%%%%%%%%%
%%%%%%%%%%%%%%---Abstract---%%%%%%%%%%%%%%%%%%%
%%%%%%%%%%%%%%%%%%%%%%%%%%%%%%%%%%%%%%%%%%%%%%%
\begin{abstract}
\vspace{0.6cm}
We investigate the dynamics of ergotropy in open systems under Markovian and non-Markovian evolutions. 
In this scenario, we begin by formulating the ergotropy of an arbitrary qubit state in terms of energy and coherence. 
Thus, we determine the conditions for ergotropy freezing and ergotropy sudden death as a consequence of the system-bath interaction. In order to use ergotropy as a resource for energy extraction in the form of work in an open-system scenario, we adopt the entropy-based formulation of quantum thermodynamics. In this approach, the work gains an additional environment-induced component, which may be present even for constant Hamiltonians. We then establish an analytical relationship between the environment-induced work and ergotropy, providing an interpretation of environment-induced work in terms of variation of ergotropy. In particular, energy transfer by environment-induced work can be performed up to a limit, which is governed by the energy cost to transit between the initial and final passive states of the quantum dynamics. We illustrate these results for qubit states evolving under non-dissipative and dissipative quantum processes.
\end{abstract}

%\pacs{03.67.Mn, 03.65.Ta, 03.65.Ud, 03.65.Wj}

\maketitle

%%%%%%%%%%%%%%%%%%%%%%%%%%%%%%%%%%%%%%%%%%%%%%%
%%%%%%%%%%%%%%---Introduction---%%%%%%%%%%%%%%%
%%%%%%%%%%%%%%%%%%%%%%%%%%%%%%%%%%%%%%%%%%%%%%%
{\section{Introduction}}
Quantum thermodynamics~\cite{Gemmer:09,Binder:18,Deffner:19} is an essential ingredient behind a full understanding of quantum technologies. 
Indeed, the engineering of quantum tasks typically entails energy transfer operations. The maximum amount 
of energy that can be extracted from a quantum system by unitary transformations is referred to as ergotropy~\cite{Allahverdyan:04}. 
Since only unitary operations are allowed, ergotropy means energy extraction in the form of work, which can be performed by an external field. On the other hand, the amount of ergotropy available for a system is also affected by its interaction with the surrounding environment. 
In this open-system picture~\cite{Breuer:Book}, heat and work may be interchanged via thermodynamic processes, allowing for the experimental realization of 
a diversity of quantum devices, such as quantum heat engines~\cite{Rossnagel:16,Ono:20,Bouton:21}, quantum refrigerators~\cite{Maslennikov:19}, quantum batteries~\cite{Joshi:22,Zhu:23}, among others. 

\begin{figure}[H]
\includegraphics[scale=0.215]{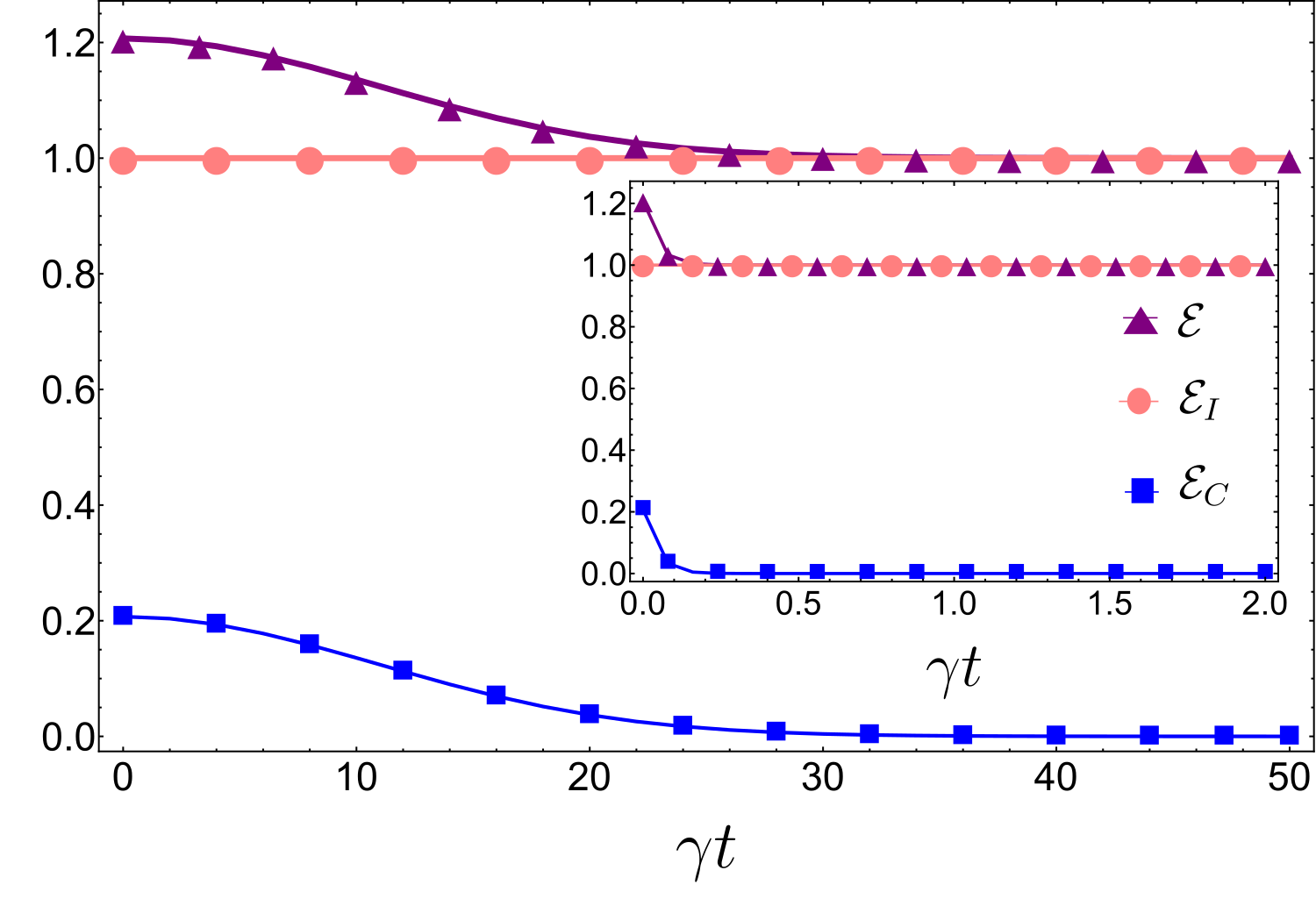}
\caption{(Color online) Dynamics of the ergotropy $\mathcal{E}$, incoherent ergotropy $\mathcal{E}_I$, and coherent ergotropy $\mathcal{E}_C$ as functions of $\gamma t$ under a non-Markovian PD channel with $\Gamma=0.01\gamma$.  Inset: Same functions under a Markovian PD channel ($\Gamma\rightarrow\infty$).  The initial conditions in both cases are $C_0=0.5$ and $U_0=0.5$.}
\label{fig:PD}
\end{figure}

The use of ergotropy as a resource for quantum tasks has to obey the energy balance dictated 
by the first law of thermodynamics, whose quantum version has been originally established by Alicki~\cite{alicki}. 
In this description, internal energy is taken as the expectation value of the underlying system Hamiltonian governing 
the quantum dynamics. Concerning work, it is associated with a change in the gap structure of the energy spectrum, 
which is induced by a time-dependent Hamiltonian. Finally, heat may be exchanged by the system as the external environment 
is taken into account, leading to a time-dependent density operator and setting a population dynamics governed by the system-bath interaction. 
Alternatively, heat may also be directly defined through the change of entropy of the quantum system~\cite{adolfo, Ahmadi:23}. 
By adopting this entropy-based formulation, the first law of thermodynamics implies that work gains an additional environment-induced 
component, which may be present even for constant Hamiltonians. 
From a foundational point of view, the entropy-based approach 
leads to the debate on   
the barrier between work and heat definitions so that the energy balance is ensured. This has implications for the analysis of 
entropy production and irreversibility for processes far from equilibrium (e.g., Ref.~\cite{Vallejo:21}). 
As an application, the environment-induced work may  
be useful for quantum devices as long as we can engineer the system-bath interaction so that the 
energy transfer from (or to) the system realizes a desired quantum task. \\
\indent In this work, we aim at exploring the environment-induced work, analytically relating it with ergotropy. We begin with the investigation of the dynamics of ergotropy in open systems under Markovian and non-Markovian evolutions. 
By expressing ergotropy for qubit states in terms of energy and coherence, we will split ergotropy in both coherent and incoherent contributions, 
obtaining the conditions for ergotropy 
sudden death and ergotropy freezing as a consequence of the system-bath interaction. These phenomena can be seen 
as analogous to the previously established sudden changes and freezing for classical and quantum correlations~\cite{Maziero:09,Paula:13}, 
including entanglement~\cite{Yu:09}. Then, we will establish a relationship between ergotropy 
variation and environment-induced work. 
More specifically, we will show that, for a constant Hamiltonian, ergotropy can be effectively changed 
by the  
environment-induced work, providing an interpretation of environment-induced work in terms of variation of ergotropy 
in the quantum system. The remaining contribution for the ergotropy variation is then given by the energy cost 
to transit between the initial and final passive 
states of the dynamics, with the passive state denoting the state with no energy available to be extracted. This result closely connects the environment-induced work with the ergotropy dynamics.\\
We illustrate our relationship by looking at qubit states evolving under an amplitude damping process. For this example, we explicitly show that the ergotropy variation is an upper limit for the environment-induced work 
throughout the dynamics. 
\\
%%%%%%%%%%%%%%%%%%%%%%%%%%%%%%%%%%%%%%%%%%%%%%%
%%%%%%%%%%---Ergotropy---%%%%%%%%%%%%%%%%%%%%%%
%%%%%%%%%%%%%%%%%%%%%%%%%%%%%%%%%%%%%%%%%%%%%%%
{\section{Ergotropy and its quantum dynamics}}
The ergotropy is defined as the maximum amount of energy that can be extracted from a quantum system via cyclic unitary operation \cite{Allahverdyan:04}, i.e., $\mathcal{E}(\rho)=\max_{V\in \mathcal{U}}\{U(\rho)-U(V\rho V^{\dagger})\}$,
where $U(\rho)=\text{tr}\left[\rho H\right]$ represents the internal energy, with $H$ and $\rho$ denoting the Hamiltonian and the density operator, respectively, and $\mathcal{U}$ the set of all unitary transformations. These transformations are required to be cyclic with respect to $H$. Assuming the spectral decomposition $\rho=\sum_nr_n\ket{r_n}\bra{r_n}$ and $H=\sum_n\varepsilon_n\ket{\varepsilon_n}\bra{\varepsilon_n}$, with the eigenstates reordered so that $r_0\ge r_1\ge ...$ and $\varepsilon_0\le \varepsilon_1\le...$, a close expression for the ergotropy can be obtained in terms of the passive state, $\rho_\pi=\sum_nr_n\ket{\varepsilon_n}\bra{\varepsilon_n}$:
\begin{equation}
\mathcal{E}(\rho)=U(\rho)-U(\rho_\pi).
\label{W}
\end{equation}
In order to explore the role played by quantum coherence, let us consider the $l_1$-norm of coherence, $C(\rho)=\min_{\delta \in \mathcal{I}}\lVert \rho - \delta \rVert_{l_1}$, with $\mathcal{I}$ representing the set of all incoherent states (i.e., diagonal states) in the basis $\{\ket{\varepsilon_n}\}$.  The minimization leads to $C(\rho)=\lVert \rho - \rho_{\Delta} \rVert_{l_1}$, where $\rho_\Delta = \sum_n\bra{\varepsilon_n}\rho\ket{\varepsilon_n}\ket{\varepsilon_n}\bra{\varepsilon_n}$ denotes the dephased state~\cite{Baumgratz:14}. In terms of $\rho_\Delta$, we can define the incoherent ergotropy, i.e., the maximum amount of energy that can be extracted from $\rho$ without altering its quantum coherence,
\begin{equation}
\mathcal{E}_I(\rho)=\mathcal{E}(\rho_\Delta),
\label{WI}
\end{equation}
as well as the coherent ergotropy,
\begin{equation}
\mathcal{E}_C(\rho)=\mathcal{E}(\rho)-\mathcal{E}(\rho_\Delta).
\label{WC}
\end{equation}
Note that $\mathcal{E}(\rho)=\mathcal{E}_I(\rho)+\mathcal{E}_C(\rho)$~\cite{Francica:20}. Now, let us consider a two-level quantum system 
governed by dimensionless Hamiltonian $H=-\sigma_z$ represented through its spectral decomposition as $H=\ket{\varepsilon_1}\bra{\varepsilon_1}-\ket{\varepsilon_0}\bra{\varepsilon_0}$, with associated eigenvalues $\varepsilon_0=-1$ and $\varepsilon_1=1$. 
In this case, quantum coherence and energy are, respectively, given by
\begin{equation}\label{Eq:coherence}
C(\rho)=2\lvert \bra{\varepsilon_0}\rho\ket{\varepsilon_1}\rvert,
\end{equation}
\begin{equation}\label{Eq:energy}
U(\rho)=1-2\bra{\varepsilon_0}\rho\ket{\varepsilon_0},
\end{equation}
where $-1\leq U \leq 1$ and $0\leq C \leq 1$ such that $U^2+C^2 \leq 1$. From Eqs.~\eqref{W}, \eqref{WI}, and~\eqref{WC}, we can express the ergotropy for an arbitrary two-level system as a function of $C$ and $U$, yielding
\begin{equation}
\mathcal{E}(C,U)=\sqrt{C^2+U^2}+U.
\label{W2}
\end{equation}
For the incoherent and coherent contributions for ergotropy, we obtain
\begin{equation}\label{Eq:WI}
\mathcal{E}_I(U)=2\max\{0,U\}=\mathcal{E}(0,U)
\end{equation}
and
\begin{equation}\label{Eq:WC}
\mathcal{E}_C(C,U)=\sqrt{C^2+U^2}-|U|=\mathcal{E}(C,-|U|), 
\end{equation}
respectively. Note that $\mathcal{E}(C,0)=\mathcal{E}_C(C,0)=C$. 
According to Eq.~\eqref{W2}, the dynamics of ergotropy will depend on the dynamics of both coherence $C$ and internal energy $U$. 
We can then determine explicit conditions for $U$ and $C$ that ensure peculiar behaviors for the ergotropy as a function of time. 
In particular, it is evident that \hypertarget{caso a}{\textcolor{blue}{\textbf{(a)}}} $\mathcal{E}(t)=\mathcal{E}_0\,\,$ iff $\,\,C(t)=C_0\,\,$ and $\,\,U(t)=U_0\,\,$\,\,$(\forall\,t)$; \hypertarget{caso b}{\textcolor{blue}{\textbf{(b)}}} $\mathcal{E}(t)=0\,\,$ iff  $\,\,C(t)=0\,\,$ and $\,\,U(t)\leq 0 \,\,$ $(\forall\,t)$, where $\mathcal{E}_0=\mathcal{E}(0)$, $C_0=C(0)$, and $U_0=U(0)$. The properties \hyperlink{caso a}{\textbf{(a)}} and \hyperlink{caso a}{\textbf{(b)}} reveal that the dynamics of ergotropy can exhibit freezing and sudden death effects, respectively. In order to explore these two phenomena, we start by defining the dynamics of a quantum system interacting with an environment in terms of Kraus representation as $\rho(t) = \sum_{i} K_i(t)\rho(0) K_i(t)^\dagger$, where the Kraus operators $\{K_i(t)\}$ satisfy the completeness relation $\sum_{i} K_{i}(t)^{\dagger} K_{i}(t) = I$ and allows the environment characterization in both Markovian and non-Markovian regimes.
\\
%%%%%%%%%%%%%%%%%%%%%%%%%%%%%%%%%%%%%%%%%%%%%%%
%%%%%%%%%%%%%%%---Freezing ---%%%%%%%%%%%%%%%%%
%%%%%%%%%%%%%%%%%%%%%%%%%%%%%%%%%%%%%%%%%%%%%%%
{\subsection{Freezing} \label{II(A)}}
According to condition \hyperlink{caso a}{{\bf (a)}}, the ergotropy freezing can be observed in non-dissipative quantum processes, such as the phase damping (PD) channel, when the coherence remains unchanged. The Kraus operators for a non-Markovian version of the PD channel are defined as \cite{TYu:10,Utagi:20}
\begin{equation}\label{Eq:PD}
    K_0(t) = \sqrt{\frac{1+e^{-q(t)}}{2}}\, I\, ,\  K_1(t) = \sqrt{\frac{1-e^{-q(t)}}{2}}\, \sigma_z\,,
\end{equation} 
 where $q(t)=\frac{\gamma}{2}\{t + \frac{1}{\Gamma}(\exp(-\Gamma t )-1) \}$. Here $\gamma^{-1}$ and $\Gamma^{-1}$  define the qubit relaxation time and the reservoir correlation time, respectively, with the Markovian regime obtained in the limit $\Gamma\rightarrow\infty$. For this non-Markovian channel, the time evolution of the coherence and energy are, respectively, given by 
\begin{equation}
    C(t)=e^{-q(t)}C_0\,\,\text{and}\,\,U(t)=U_0. 
\end{equation}
Thus, from Eq.~(\ref{W2}), a freezing ergotropy at a nonvanishing value $2U_0$ is accomplished for zero initial coherence and positive initial energy, i.e.,
\begin{equation}\label{Eq: Ergotropy}
\mathcal{E}(t)=2U_0\,\,\,\forall\, t \,\,\, \text{iff} \,\,\,C_0=0\,\,\,\text{and}\,\,\, U_0>0.
\end{equation}
As a consequence of Eqs.~(\ref{Eq:WI}) and~(\ref{Eq: Ergotropy}), $\mathcal{E}_{I}(t)=2U_0$  $\forall \, t$ iff $U_0>0$ (for all $C_0$). To illustrate this phenomenon, we show $\mathcal{E}$, $\mathcal{E}_I$, and $\mathcal{E}_C$ in Fig.~\ref{fig:PD} as functions of $\gamma t$ for $C_0=0.5$ and $U_0=0.5$, where the initial state is taken as $\rho_{00}=0.25$, $\rho_{11}=1-\rho_{00}$, and $|\rho_{01}|=0.25$. In both non-Markovian and Markovian regimes, notice that the incoherent component freezes for all $\gamma t$, while the coherent part exhibits a monotonic decay. In terms of the total ergotropy, the contribution of its coherent component allows the ergotropy to decay until it achieves a steady state. Additionally, it is clear to see in Fig.~\ref{fig:PD} a delayed ergotropy decay in the non-Markovian regime in comparison with the Markovian regime. 

%%%%%%%%%%%%%%%%%%%%%%%%%%%%%%%%%%%%%%%%%%%%%%%
%%%%%%%%%%%%---Sudden death---%%%%%%%%%%%%%%%%%
%%%%%%%%%%%%%%%%%%%%%%%%%%%%%%%%%%%%%%%%%%%%%%%
{\subsection{Sudden death}\label{II(B)}}
Condition \hyperlink{caso b}{\textbf{(b)}} ensures that the sudden death phenomenon can be achieved in dissipative quantum processes, such as the amplitude damping (AD) channel, when coherence is absent. The Kraus operators for a non-Markovian AD channel are given by \cite{Bellomo:07} 
\begin{equation}
    K_0(t)=\left[
\begin{array}{cc}
1 & 0 \\
0 & \sqrt{q(t)} 
\end{array} \right],\
 \, \, K_1(t)=\left[
\begin{array}{cc}
0 & \sqrt{1-q(t)} \\
0 & 0 
\end{array} \right],\
\label{ADsd}
\end{equation}

where $q(t)=\exp({-\Gamma t})\{\cos(\frac{dt}{2})+\frac{\Gamma}{d}\sin(\frac{dt}{2})\}^2$ with $d=\sqrt{2\gamma\Gamma -\Gamma^2}$. The spectral width $\Gamma$ and the coupling strength $\gamma$ are related to the reservoir correlation time ($\Gamma^{-1}$) and the qubit relaxation time ($\gamma^{-1}$), respectively.  The dynamics of coherence and energy for this non-Markovian channel are, respectively, determined by 
\begin{equation}
    C(t)=\sqrt{q(t)}C_0\,\,\text{and} \,\,U(t)=(1+U_0)q(t)-1. 
\end{equation}
As $U(t)<U_0$ for all $t>0$, ergotropy collapses and revivals can be observed if the initial coherence is zero and the initial energy is positive, in agreement with the property \textbf{(b)}. These sudden changes occur when the energy changes its sign during the quantum process. Consequently, the sudden change times $\{t_n\}$ satisfy the condition
\begin{equation}
    q(t_n)=\frac{1}{1+U_0}\,\,\,\,\text{with}\,\,\,\,t_1<t_2<...<t_{sd},
\end{equation}
where odd and even values of $n$ indicate sudden deaths and births, respectively, with $t_{sd}$ characterizing an eternal death. Thus, we conclude that
\begin{equation}
\mathcal{E}(t)=0\,\,\,\forall\, t\geq t_{sd} \,\,\, \text{iff} \,\,\,C_0=0\,\,\, \text{and}\,\,\, U_0>0.
\end{equation}
Consequently, $\mathcal{E}_I(t)=0$ $\forall\, t\geq t_{sd}$ iff $U_0>0$ (for all $C_0$). In the Markovian regime, $\Gamma\rightarrow\infty$, the emergence of eternal death is 
\begin{figure}[H]
\includegraphics[scale=0.22]{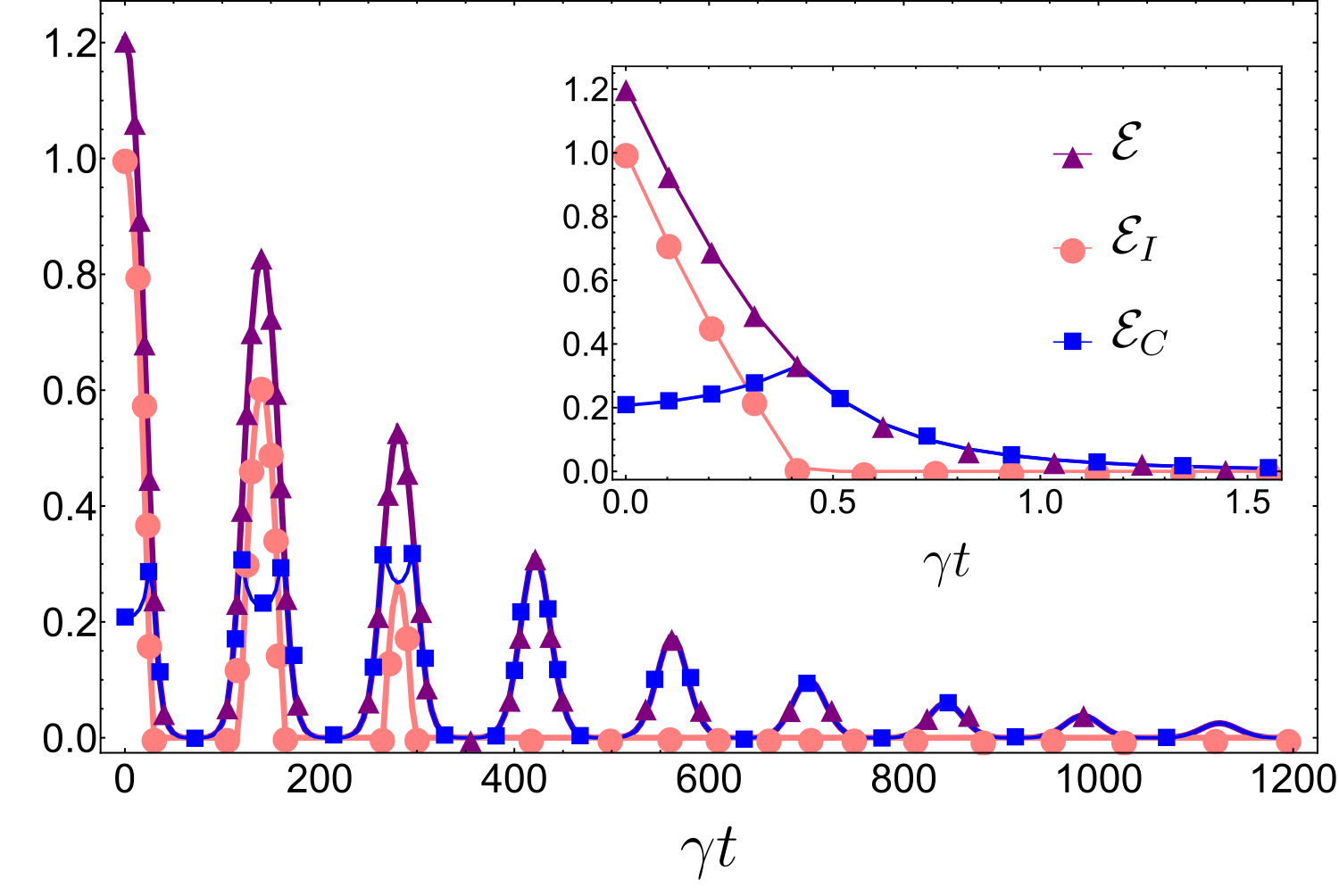}
\caption{(Color online) 
 Dynamics of the ergotropy $\mathcal{E}$, incoherent ergotropy $\mathcal{E}_I$, and coherent ergotropy $\mathcal{E}_C$ as functions of $\gamma t$ under a non-Markovian AD channel with $\Gamma=0.001\gamma$.  Inset: Same functions under a Markovian AD channel ($\Gamma \rightarrow \infty$).  The initial conditions in both cases are $C_0=0.5$ and $U_0=0.5$.}
\label{fig:AD}
\end{figure}

\noindent determined by $t_{sd}=t_1=\gamma^{-1}\ln(1+U_0)$. In this limit, there are no temporary collapses and revivals.

Figure~\ref{fig:AD} shows the ergotropy, as well as its incoherent and coherent parts, as functions of $\gamma t$ for the initial conditions $C_0 = 0.5$, $U_0 = 0.5$, and initial state given by $\rho_{00}=0.25$, 
$\rho_{11}=1-\rho_{00}$, and $|\rho_{01}|=0.25$. Remarkably, the incoherent component exhibits a non-monotonic (monotonic) decay until the eternal death time $t_{sd} \approx 297/\gamma$ ($t_{sd} \approx 0.405/\gamma$) for the non-Markovian (Markovian) regime, with $\Gamma=0.001\gamma$ ($\Gamma \rightarrow \infty$). Besides, notice that the coherent part contributes to the inhibition of this phenomenon. 
\\
\\

%%%%%%%%%%%%%%%%%%%%%%%%%%%%%%%%%%%%%%%%%%%%%%%
%%%%%%%%%---Environment-induced work---%%%%%%%%
%%%%%%%%%%%%%%%%%%%%%%%%%%%%%%%%%%%%%%%%%%%%%%%
{\section{Environment-induced work}}
In order to use ergotropy as a resource, we will investigate how to explicitly extract energy in the form of work via the system-environment interaction. 
The energy balance in a thermodynamic process is ruled by the first law of thermodynamics. In its quantum version, a standard formulation can be written 
as~\cite{alicki}
\begin{equation}
dU = \delta \Tilde{Q}  + \delta \Tilde{W},
\end{equation}
where $dU$ is the infinitesimal internal energy change, 
$\delta \Tilde{Q}$ is the infinitesimal heat exchanged in the process, 
and $\delta \Tilde{W}$ is the infinitesimal work performed by (or on) the system, 
with $U = \text{Tr} (H \rho)$, $\delta \Tilde{Q} = \text{Tr} (H \, d\rho)$, and $\delta \Tilde{W} = \text{Tr} (dH \, \rho)$. 
As an alternative formulation, we can modify the definition of $\delta \Tilde{Q}$ so that heat is directly linked with the entropy variation. 
In this entropy-based framework for quantum thermodynamics~\cite{adolfo}, heat and work are defined through 
$\delta Q=\delta \Tilde{Q}- \delta W^*$ and $\delta W=\delta \Tilde{W}+\delta W^*$, so that 
the first law of thermodynamics is
\begin{equation} \label{Eq:1-law}
    dU = \delta Q +\delta W,  
\end{equation}
with $\delta W^*$ introduced as an environment-induced work~\cite{adolfo}
\begin{equation}\label{Eq:work*}
\delta W^*= \sum_n r_n \left(\left\langle r_n\right|H\,d\left|r_n\right\rangle+h.c.\right),
\end{equation}
where $\ket{r_n}$ represents an eigenvector of the density operator $\rho$ and $r_n$ the corresponding eigenvalue.
Notice that the first law of thermodynamics is preserved, with the internal energy infinitesimal $dU$ kept unchanged 
due to the new definitions of heat and work. From this point on, unless stated otherwise, 
heat and work will refer to the entropy-based formulation of quantum thermodynamics.

A utility for the definition of heat $Q$ as a witness of non-Markovianity for unital quantum maps has been recently provided~\cite{john}. Now, we will provide an operational meaning for $W^*$ in terms of ergotropy variation.  
We consider a quantum system described by an initial density operator $\rho(t=0) \to \rho_0 $ and governed by a constant Hamiltonian $H$. In this scenario, the conventional work $\Tilde{W}$ is null. The system interacts with an external environment and is taken to a final density operator $\rho(t_c) \to \rho_c$ at a specific characteristic time $t_c$ such that 
the total heat $Q$ exchanged with the environment is vanishing. For this effective adiabatic process, the environment-induced work is the only contribution to the energy balance 
in the first law of thermodynamics, i.e.,
\begin{equation}
Q(\rho_c)=0
\end{equation}
and, consequently,
\begin{equation}
W^{*}(\rho_c)=\Delta U(\rho_c)=U(\rho_c)-U(\rho_0).
\label{Energ_var}
\end{equation}
By examining the ergotropy variation,
\begin{equation}
\Delta \mathcal{E}(\rho_c)= \mathcal{E}(\rho_c)- \mathcal{E}(\rho_0),
\label{Erg_var}
\end{equation}
by using Eq.~(\ref{W}), we can write
\begin{equation}
\Delta \mathcal{E}(\rho_c)= [U(\rho_c)-U(\rho_{c\pi})] - [U(\rho_0)-U(\rho_{0\pi})],
\label{Erg_var2}
\end{equation}
where $\rho_{c\pi}$ and $\rho_{0\pi}$ are the passive states associated with $\rho_c$ and $\rho_0$, respectively. Finally, by using Eq.~(\ref{Energ_var}), and defining the passive energy variation,
\begin{equation}
\Delta U_\pi(\rho_c) = U(\rho_{c\pi})- U(\rho_{0\pi}),
\end{equation}
we obtain
\begin{equation}
\Delta \mathcal{E}(\rho_c)= W^*(\rho_c) - \Delta U_\pi(\rho_c).
\label{Erg_var3}
\end{equation}
The contribution $\Delta U_\pi(\rho_c)$ for the ergotropy can be interpreted as the energy cost to transit between the initial and final passive states $\rho_{0\pi}$ and $\rho_{c\pi}$, respectively. We observe that Eq.~(\ref{Erg_var3}) agrees with the discussion about the energetics of the ergotropy 
in Ref.~\cite{binder}, with $\Delta U_{\pi}$ defined there as an operational heat. Here, we can then directly connect the 
environment-induced work $W^*$ with the variation of ergotropy for constant Hamiltonians, reinforcing the interpretation of $W^*$ as an effective work extracted due to the system-environment interaction. For 2-level systems, the Hamiltonian is given by $H=-\sigma_z$ %($\omega_0$ is the energy gap) %
and the density matrix, in terms of the Bloch sphere, can be written as $\rho(t)=\left(I+\vec{r}(t)\cdot\vec{\sigma}\right)/2$, where $\vec{r}(t)=[x(t),y(t),z(t)]$ is the Bloch vector and $I,\Vec{\sigma}$ are Pauli operators. In this case, coherence and internal energy are given by $C(t)=\sqrt{x(t)^2+y(t)^2}$ and $U(t)=- z(t)$, respectively. Since $\delta Q =(U/r)dr$ ~\cite{john}, the 
\begin{figure}[t]
		\centering		\includegraphics[scale=0.4]{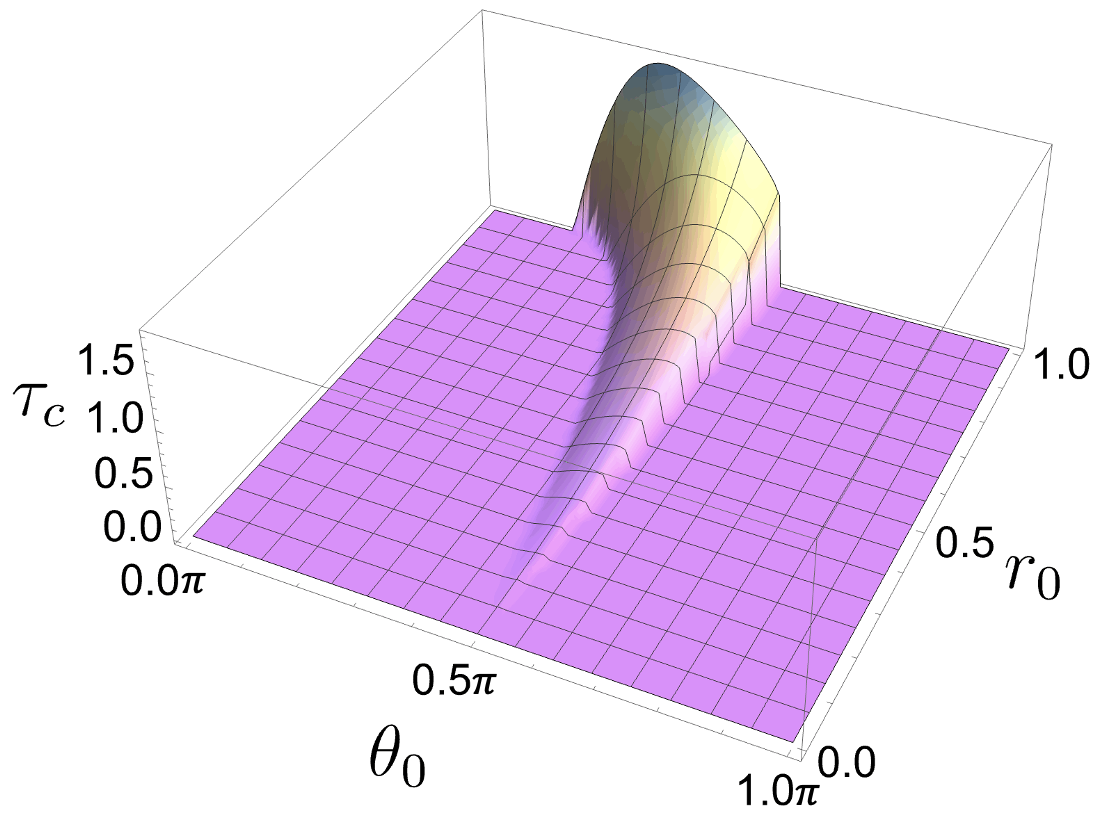}
  \caption{(Color online) Dimensionless characteristic time $\tau_c$ as a function of the initial state $(r_0,\,\theta_0)$ for $0\leq r_0\leq 1$ and $0\leq \theta_0 \leq \pi$.}
  \label{fig:tc}
	\end{figure}
characteristic %(thermal) 
adiabatic time $t_c$ can be obtained through the condition
\begin{equation}
Q(t_c)=-\int_{0}^{t_c}\frac{z(t)}{r(t)}\dot{r}(t) \,dt=0.
\label{Eq:Qzero}
\end{equation}
For the components of $\Delta\mathcal{E}(t_c)$ in Eq.~(\ref{Erg_var3}), 
we have
\begin{equation}
 W^*(t_c)= -\Delta z(t_c),
\end{equation}
\begin{equation}
 \Delta U_{\pi}(t_c)= -\Delta r(t_c).
\end{equation}
Thus, we conclude that the work $W^*(t_c)$ and the passive energy variation $-\Delta U_{\pi}(t_c)$ are associated with the ergotropic cost of rotation and scale transformation 
(dilation or contraction) of the Bloch vector $\vec{r}$, respectively. According to Eq.~(\ref{Eq:WI}), the incoherent ergotropic variation, $\Delta\mathcal{E}_I$, vanishes for quantum processes with constant or non-positive energy. In these cases, the ergotropy variation is purely coherent, i.e., $\Delta \mathcal{E}(t)=\Delta \mathcal{E}_C(t)$ if $\dot{U}(t)=0$ or $U(t)\leq 0$ for all $t$. For non-dissipative quantum processes,
\begin{equation}
\Delta \mathcal{E}_C(t_c)= W^*(t_c)=\Delta U_\pi(t_c)=0\,\,\,(\dot{U}=0).
\end{equation}
Thus, the PD process discussed in Sec. \ref{II(A)} is unable to extract the available quantum resource $\mathcal{E}_C$ through environment-induced work. In other words, there is no effective adiabatic process with $W^*(t_c)\neq0$ for PD channels. On the other hand, the extraction is possible for dissipative quantum processes such as the AD channel described in Sec. \ref{II(B)}.

Here, we will illustrate the connection between $W^*$ and $\Delta \mathcal{E}$ by considering the paradigmatic model of the decay of an excited state of a 2-level atom interacting with an environment by spontaneous emission~\cite{Nielsen-Chuang, Breuer:09} (a Markovian AD process).  The spontaneous emission process is governed by the Markovian master equation
\begin{equation} \label{Eq:mastereq-markov}
        \dot{\rho}(t)=i\left[\sigma_z, \rho(t)\right]+\gamma
    \left[\sigma^{-}\rho(t)\sigma^{+}-\frac{1}{2}\left\{\sigma^{+}\sigma^{-},\rho(t)\right\}\right],
\end{equation}

\begin{figure}[t]
		\centering
		\includegraphics[scale=0.42]{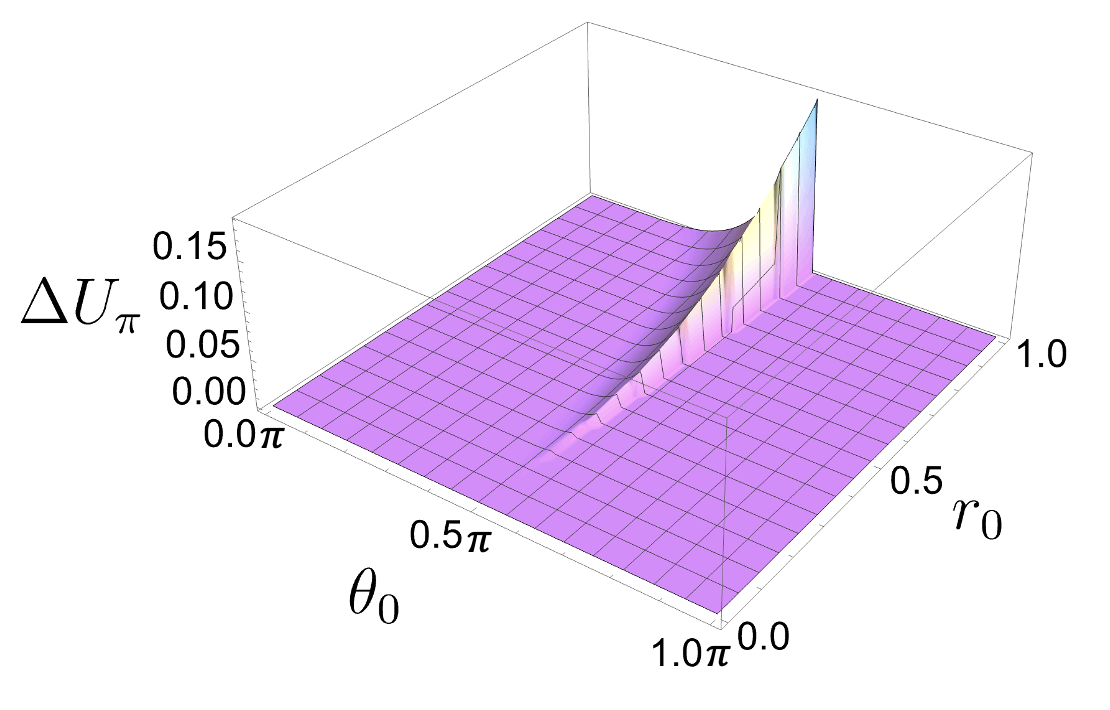}
  \caption{(Color online) Passive energy cost $\Delta U_\pi$ as a function of the initial state $(r_0,\,\theta_0)$ for $0\leq r_0\leq 1$ and $0\leq \theta_0 \leq \pi$.}
  \label{fig:DeltaUpi}
	\end{figure}
	
\noindent where $\gamma$ is the dissipation rate of spontaneous emission and 
$\sigma^{+}=(\sigma_x -i \sigma_y)/2$ and $\sigma^{-}=(\sigma_x +i \sigma_y)/2$
are the raising and lowering operators for a 2-level atom. 
Notice that the ground state of $H$ is the computational state $|0\rangle$ in the north pole of the Bloch sphere, which is the expected long time limit after energy loss in the spontaneous 
emission dynamics. 
The solution of Eq. (\ref{Eq:mastereq-markov}) with an arbitrary initial state $\Vec{r}_0=[x_0,y_0,z_0]$ is given by $\Vec{r}(t)=[x(t),y(t),z(t)]$ with
\begin{equation}\label{Eq:rx}
x(t) =e^{-\gamma t/2}\left[x_0\cos{2 t} + y_0\sin{2 t} \right], 
\end{equation}
\begin{equation} \label{Eq:ry}
y(t) =e^{-\gamma t/2}\left[y_0\cos{2 t} -x_0\sin{2 t} \right],
\end{equation}
\begin{equation} \label{Eq:rz}
z(t) =e^{-\gamma t}\left[-1+z_0+ e^{\gamma t} \right].
\end{equation}
According to this solution and Eq.~(\ref{Eq:Qzero}), the dimensionless characteristic adiabatic time $\tau_c=\gamma t_c$ is a function only of the initial parameters $r_0=(x_0^2+y_0^2+z_0^2)^{1/2}$ and $\theta_0=\text{arccos}[z_0/r_0]$, where $0\leq r_0\leq 1$ and $0\leq \theta_0 \leq \pi$. We numerically investigate

\begin{figure}[H]
\includegraphics[scale=0.305]{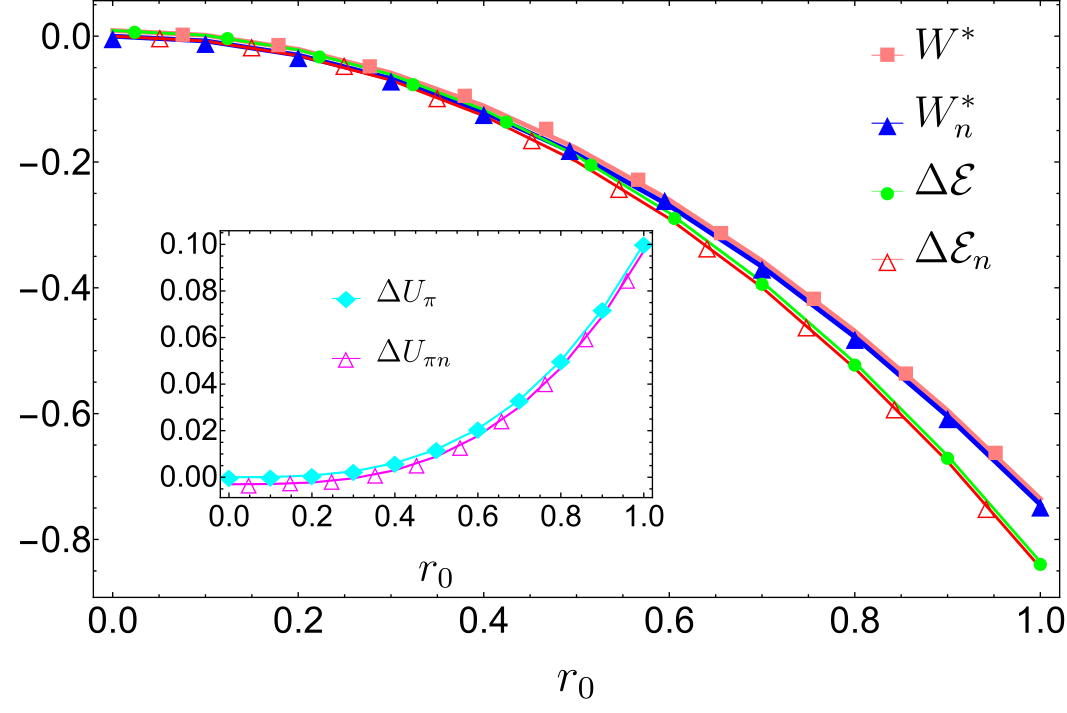}
\caption{(Color online) Environment-induced
work in the Markovian $W^*$ and non-Markovian $W^{*}_{n}$ regimes, as well as ergotropy variation in the Markovian $\Delta \mathcal{E}$ and non-Markovian $\Delta \mathcal{E}_{n}$ regimes, as functions of $r_0$ ($0\leq r_0 \leq 1$) for $\theta_0 = \pi/2$. For the non-Markovian dynamics, we adopted $\Gamma=0.01 \gamma$. Inset: Passive energy cost for both Markovian $\Delta U_{\pi}$ and non-Markovian $\Delta U_{\pi n}$ regimes.}
\label{fig:fammix}
\end{figure}

\begin{figure}[H]
\includegraphics[scale=0.1905]{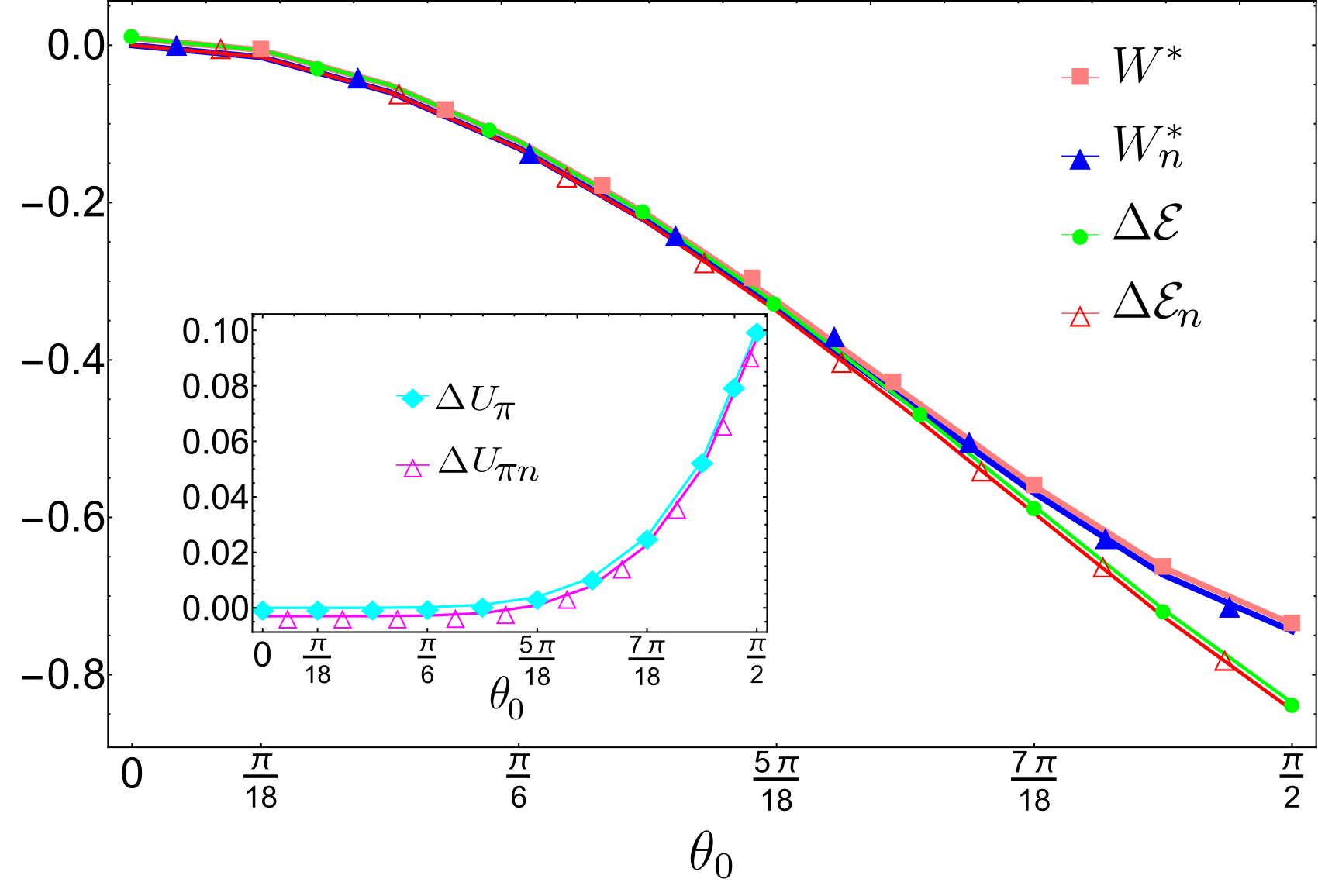}
\caption{(Color online)  Environment-induced
work in the Markovian $W^*$ and non-Markovian $W^{*}_{n}$ regimes, as well as ergotropy variation in the Markovian $\Delta \mathcal{E}$ and non-Markovian $\Delta \mathcal{E}_{n}$ regimes, as functions of $\theta_0$ ($0\leq \theta_0 \leq \pi/2$) for $r_0 = 1$. For the non-Markovian dynamics, we adopted $\Gamma=0.01 \gamma$. Inset: Passive energy cost for both Markovian $\Delta U_{\pi}$ and non-Markovian $\Delta U_{\pi n}$ regimes.}
\label{fig:pure}
\end{figure}
\vspace{1pt}
\noindent the characteristic adiabatic time for arbitrary initial states $\vec{r}_0=(r_0,\theta_0)$ in Fig. \ref{fig:tc}. Notice that $\tau_c$ is not negligible for initial states close to $\theta_0=\pi/2$  or $r_0=1$, mainly for $\theta_0\geq \pi/2$ (north hemisphere), where $\Delta \mathcal{E}=\Delta \mathcal{E}_C$. We also study the energy cost $\Delta U_\pi$ as a function of $r_0$ and $\theta_0$. This is exhibited in Fig.~\ref{fig:DeltaUpi}. Notice that $\Delta U_\pi\geq 0$ for all initial states. Therefore, from Eq.~(\ref{Erg_var3}), the work $W^*(t_c)$ performed on the system by the environment is not greater (in absolute value) than the ergotropy variation $\Delta \mathcal{E}(t_c)$. This result implies that the environment cannot provide more energy to the system than it can be extracted via the definition of extractable work through a variation of ergotropy. Finally, we plot $\Delta \mathcal{E}(t_c)$ and $W^*(t_c)$ for two particular types of families of initial states: a family of mixed states located on the equatorial plane of the Bloch sphere (see Fig.~\ref{fig:fammix}) and a family of pure states located on the upper surface of the Bloch sphere (see Fig.~\ref{fig:pure}), where $t_c\neq 0$. As a by-product, we can also determine the energy cost associated with the remaining contribution $\Delta U_\pi(t_c)$ for the ergotropy variation, as shown in the inset of Fig.~\ref{fig:fammix} and Fig.~\ref{fig:pure}. In theses cases, it is evident that $W^*(t_c)/ \Delta \mathcal{E}(t_c)\leq 1$, where $\Delta \mathcal{E}(t_c)=\Delta \mathcal{E}_C(t_c)$. Notice also that the singular behavior of the dimensionless characteristic time shown in Fig. 3 is not manifested in Figs.~\ref{fig:fammix} and~\ref{fig:pure}. This occurs because we have kept the dynamics in the north hemisphere of the Bloch sphere, with $\theta_0 \le \pi/2$.

We can also consider a more general dynamics, such as the non-Markovian case. This can be analyzed using the physical process as in Eq.~(\ref{ADsd}), from which Eq.~(\ref{Eq:mastereq-markov}) follows as a Markovian limit. For each initial state, instead of only a single $t_{c}$, a set of characteristic times emerges, which is denoted by $\{t_{nc}\}$. By looking at these characteristic times,
we show that, by taking the largest $t_{nc}$ for each state, we obtain results very close to the Markovian case,  
as shown in Fig.~\ref{fig:fammix} and 
Fig.~\ref{fig:pure}. This behavior holds independently of the ratio $\Gamma/\gamma$.

\section{Conclusions}
We have investigated the dynamics of ergotropy and the energy extraction of quantum systems via environment-induced work $W^*$
in the entropy-based formulation of quantum thermodynamics. By introducing an analytical expression for $W^*$ in terms of the 
ergotropy variation $\Delta \mathcal{E}$ under a generic quantum map, we have explicitly found $W^*$ 
as a resource for system-environment energy transfer. Indeed, by considering a constant system Hamiltonian and by assuming the engineering of the 
system-environment interaction, we have explicitly shown that $W^*$ is able to induce energy transfer up to a fixed limit in a qubit system under amplitude damping, 
with the remaining contribution for $\Delta \mathcal{E}$ provided by the energy cost 
to transit between the initial and final passive states. This motivates further considerations, such as the investigation of this passive energy cost in other open-system evolutions. Moreover, additional analysis can also be fruitful for the contribution of $W^*$ for the efficiency 
of quantum heat engines in the machine steps where contact with the thermal bath occurs. These considerations in the view of the first and second laws 
of thermodynamics are potential targets of future developments. 

{\section*{Acknowledgments}} 
J.M.Z.C. acknowledges Conselho Nacional de Desenvolvimento Cient\'{\i}fico e Tecnol\'ogico (CNPq) for financial support. 
M.S.S. is supported by Conselho Nacional de Desenvolvimento Cient\'{\i}fico e Tecnol\'ogico (CNPq) (307854/2020-5). 
This research is also supported in part by Coordena\c{c}\~ao de Aperfei\c{c}oamento de Pessoal de N\'{\i}vel Superior (CAPES) (Finance Code 001) 
and by the Brazilian National Institute for Science and Technology of Quantum Information (INCT-IQ).


\begin{thebibliography}{30}

\bibitem{Gemmer:09} J. Gemmer, M. Michel, and G. Mahler, {\it{Quantum Thermodynamics}}, (Berlin:Springer, 2009).

\bibitem{Binder:18} F. Binder, L. A. Correa, G. Gogolin, J. Anders, and G. Adesso, {\it{Thermodynamics in the Quantum Regime. Fundamental Theories of Physics}}, 
(Berlin:Springer, 2018).

\bibitem{Deffner:19} S. Deffner and S. Campbell, {\it{Quantum Thermodynamics: An introduction to the thermodynamics of quantum information}}, 
(Morgan \& Claypool Publishers, 2019).

\bibitem{Allahverdyan:04} A. E. Allahverdyan, R. Balian, and Th. M. Nieuwenhuizen, Europhys. Lett. {\bf 67}, 565 (2004).

\bibitem{Breuer:Book} H.-P. Breuer and F. Petruccione, {\it{The Theory of Open Quantum Systems}}, Oxford University Press, USA, 2007.

\bibitem{Rossnagel:16} J. Rossnagel, S. T. Dawkins, K. N. Tolazzi, O. Abah, E. Lutz, F. Schmidt-Kaler, and K. Singer, 
Science {\bf 352}, 325 (2016).

\bibitem{Ono:20} K. Ono, S. N. Shevchenko, T. Mori, S. Moriyama, and F. Nori, Phys. Rev. Lett. 125, 166802 (2020).

\bibitem{Bouton:21} Q. Bouton, J. Nettersheim, S. Burgardt, D. Adam, E. Lutz, and A. Widera, Nat. Commun. {\bf 12}, 2063 (2021). 

\bibitem{Maslennikov:19} G. Maslennikov, S. Ding, R. Hablützel, J. Gan, A. Roulet, S. Nimmricher, J. Dai, V.Scarani, and D. Matsukevic {\it{et al.}}, Nat. Commun. {\bf 10}, 202 (2019).

\bibitem{Joshi:22} J. Joshi and T. S. Mahesh, Phys. Rev. A {\bf 106}, 042601 (2022).

\bibitem{Zhu:23} G. Zhu, Y. Chen, Y. Hasegawa, and P. Xue, Phys. Rev. Lett. {\bf 131}, 240401 (2023).

\bibitem{alicki} R. Alicki, J. Phys. A {\bf 12}, L103 (1979).

\bibitem{adolfo} S. Alipour, A. T. Rezakhani, A. Chenu, A. del Campo, and T. Ala-Nissila , Phys. Rev. A {\bf 105}, L040201 (2022)

\bibitem{Ahmadi:23} B. Ahmadi, S. Salimi, A. S. Khorashad, Sci. Rep. {\bf 13}, 160 (2023).

\bibitem{Vallejo:21} A. Vallejo , A. Romanelli, and R. Donangelo, Phys. Rev. E {\bf 103}, 042105 (2021).

\bibitem{Maziero:09}  J. Maziero, L. C. Celeri, R. M. Serra, and V. Vedral, Phys. Rev. A {\bf 80}, 044102 (2009).

\bibitem{Paula:13} F. M. Paula, I. A. Silva, J. D. Montealegre, A. M. Souza, E. R. deAzevedo, R. S. Sarthour, A.Saguia, I. S. Oliveira, D. O. Soares-Pinto, G. Adeso {\it{et al.}}, Phys. Rev. Lett. {\bf 111}, 250401 (2013).

\bibitem{Yu:09} T. Yu and J. H. Eberly, Science {\bf 323}, 598 (2009).

\bibitem{Baumgratz:14} T. Baumgratz, M. Cramer, and M.B. Plenio, Phys. Rev. Lett. {\bf 113}, 140401 (2014).

\bibitem{Francica:20} G. Francica, F. C. Binder, G. Guarnieri, M.T. Mitchison,  J. Goold,  and F. Plastina, Phys. Rev. Lett. {\bf 125}, 180603 (2020).

\bibitem{Utagi:20} S. Utagi, R. Srikanth, and  S. Banerjee,  Sci Rep {\bf 10}, 15049 (2020). 

\bibitem{TYu:10} T. Yu, and J.H. Eberly, 
Optics Communications {\bf 283}, 676 (2010).

\bibitem{Bellomo:07}  B. Bellomo, R. Lo Franco,  and  G. Compagno, Phys. Rev. Lett. {\bf 99}, 160502, (2007).

\bibitem{john} J.~M.~Z.~Choquehuanca, F.~M.~de Paula and M.~S.~Sarandy, Phys. Rev. A \textbf{107}, 012220 (2023).

\bibitem{binder} F. C. Binder, S. Vinjanampathy, K. Modi and J. Goold,
New J. Phys. {\bf 17}, 075015 (2015).

\bibitem{Nielsen-Chuang} M. A. Nielsen and I. L. Chuang, \textit{Quantum Computation and Quantum Information}(Cambridge University Press, Cambridge, 2000).

\bibitem{Breuer:09} H.-P. Breuer, E.-M. Laine, and J. Piilo, Phys. Rev. Lett. {\bf 103},
210401 (2009).

\end{thebibliography}
\end{document}